%% file: main.tex
\documentclass{article} 
\usepackage{geometry}

\input{math_commands.tex}

\usepackage{hyperref}
\usepackage{url}
\usepackage{graphicx}
\usepackage{xcolor}

\title{Identifying DNA Sequence Motifs Using Deep Learning}

\author{Asmita Poddar$^*$, Vladimir Uzun$^\dag$, Elizabeth Tunbridge$^*$, \\ Wilfried Haerty$^\dag$, Alejo Nevado-Holgado$^*$ \\
\\
\small{$^*$University of Oxford, $^\dag$Earlham Institute}
}
\date{July 2022}

\begin{document}

\maketitle

\begin{abstract}
Splice sites play a crucial role in gene expression, and accurate prediction of these sites in DNA sequences is essential for diagnosing and treating genetic disorders. We address the challenge of splice site prediction by introducing DeepDeCode, an attention-based deep learning sequence model to capture the long-term dependencies in the nucleotides in DNA sequences. 
We further propose using visualization techniques for accurate identification of sequence motifs, which enhance the interpretability and trustworthiness of DeepDeCode.
We compare DeepDeCode to other state-of-the-art methods for splice site prediction and demonstrate its accuracy, explainability and efficiency. Given the results of our methodology, we expect that it can used for healthcare applications to reason about genomic processes and be extended to discover new splice sites and genomic regulatory elements.
\end{abstract}

\section{Introduction}

Advancements in genomic technologies \cite{highthroughputseq} have enabled the generation of vast amounts of DNA sequence data, enabling the structural and functional study of the human genome \cite{bioinformatics}. This has been valuable in providing insights into the genetic basis of diseases. Since splice sites play a crucial role in this, accurately predicting these sites in DNA sequences is essential for diagnosing diseases.

\begin{figure}[h!]
\centering
\includegraphics[width=0.4\textwidth]{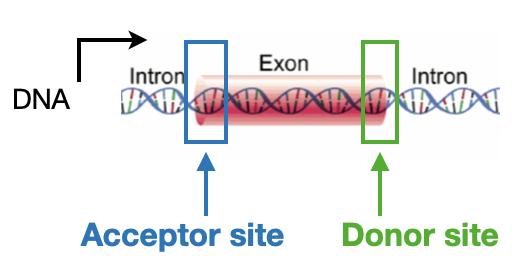}
\caption{DNA sequence with alternating introns and exons. We show the acceptor and donor splice sites within the sequence.}
\label{fig:intron-exon}
\end{figure}

The internal structure of DNA sequences consists of alternating protein-coding regions that contain information for the production of proteins called \textbf{exons}, and non-coding regions that interrupt the coding sequence called \textbf{introns}, as shown in Figure \ref{fig:intron-exon}. Mutations in the intronic region have been associated with developmental disorders such as isolated Pierre Robin sequence \cite{pierre-robin} and several types of cancer \cite{cancer}.
Hence, the accurate identification of boundaries between the exons and introns, known as \textbf{splice sites}, has special biological significance with healthcare implications, such as to study the sources of unresolved genetic disease. 

Current annotations of the human genome to identify where exons/introns are located in the sequence \cite{genomics-scope} are far from complete and much remains unknown about the DNA sequence motifs that indicate the presence of a splice site. We aim to address this challenge computationally by predicting the presence of splice sites --- both the exon start (intron end) sites, known as \textbf{acceptor sites}, and the exon end (intron start) sites, known as \textbf{donor sites} --- within the DNA sequence. 

The sequential nature of genomic data as well as the common syntactical combinatorial structure between language and the genetic code \cite{genelinguistics} makes it suitable for natural language processing architectures to be applied to the genomics domain. In this work, we propose a novel deep neural network-based methodology called DeepDeCode for trustworthy prediction of splice sites.
Typically, most existing methods use limited datasets \cite{primary-sequence-splicing-dl, splicerover}, however, to ensure generalizability, we use DNA sequences obtained from the entire human genome. 

Current methods also have limited model interpretability \cite{dna_level_splice_junc, deepsplice}. However, machine learning models used in healthcare must be trustworthy and transparent to ensure their safe and ethical use. 
To address this, we use various intrinsic and extrinsic visualization techniques to ensure that our methods and results can be confidently used to make downstream healthcare inferences. 
We expect that our method can be extended for the discovery of other structural genomic patterns that were not previously located through traditional wet lab methods, which has the potential to improve the diagnosis and treatment of genetic disorders.

\section{Related Work}

In recent years, sequence-based biological tasks such as gene expression regulation \cite{gene-regulation} and protein structure prediction \cite{protein-struct-pred} have been addressed by deep learning techniques. Motivated by this, deep learning has also been employed for splice site detection from genomic data. 
Models like SpliceVec \cite{splicevec} use multi-layer perceptrons; DanQ \cite{danq}, SpliceRover \cite{splicerover}, SpliceFinder \cite{splicefinder} and Pangolin \cite{pangolin} use convolutional neural networks (CNNs); while \cite{dna_level_splice_junc} use recurrent neural networks (RNNs). However, CNNs do not capture long term sequence dependencies and RNNs suffer from the vanishing gradient problem. 

To overcome these drawbacks, long short-term memory (LSTM)\cite{lstm} based approaches have been developed \cite{bilstm-rnn}, however these works do not investigate model explainability which is critical in biology \cite{interpretability} --- a key emphasis in our work is the explainability of our model.
Recently, to process sequences piecewise concurrently, \cite{DNABERT} developed DNABERT which is based on transformers \cite{bert} and performs similarly to LSTMs, however, their computational cost can be prohibitive (DNABERT's pretraining takes over 25 days using 8 NVIDIA 2080 Ti GPUs). In this work, we combine their attention mechanism along with LSTMs, which process data sequentially and are well-suited for sequential data such as DNA.

Furthermore, while the majority of the methods mentioned above mainly consider sequences with canonical splice sites, in this work we additionally identify non-canonical splice sites which can be harder to detect.\footnote{Canonical splice sites are those that are most commonly found in the genome (in about $99\%$ of introns). They have the nucleotides $GT$ at the donor site and $AG$ at the acceptor site.}

\section{DeepDeCode Model} \label{method}

The splice site prediction is
formulated as a binary classification task whose output represents whether a splice site in a DNA sequence of length $L$ is present ($1$) or not ($0$). 
We assume our training set $D$ contains $N$ labeled pairs.

Sequence to sequence modeling  uses an encoder-decoder architecture, where the model represents the entire input sequence using an encoder, and then decodes the encoded representation in order to predict the target.
Taking inspiration from the successful implementation and improved interpretability of seq2seq models \cite{attend&predict}, the proposed DeepDeCode model uses a bidirectional LSTM (Bi-LSTM) encoder and multi-layer perceptron decoder.

\begin{figure}[h!]
\begin{center}
\includegraphics[width=0.7\textwidth]{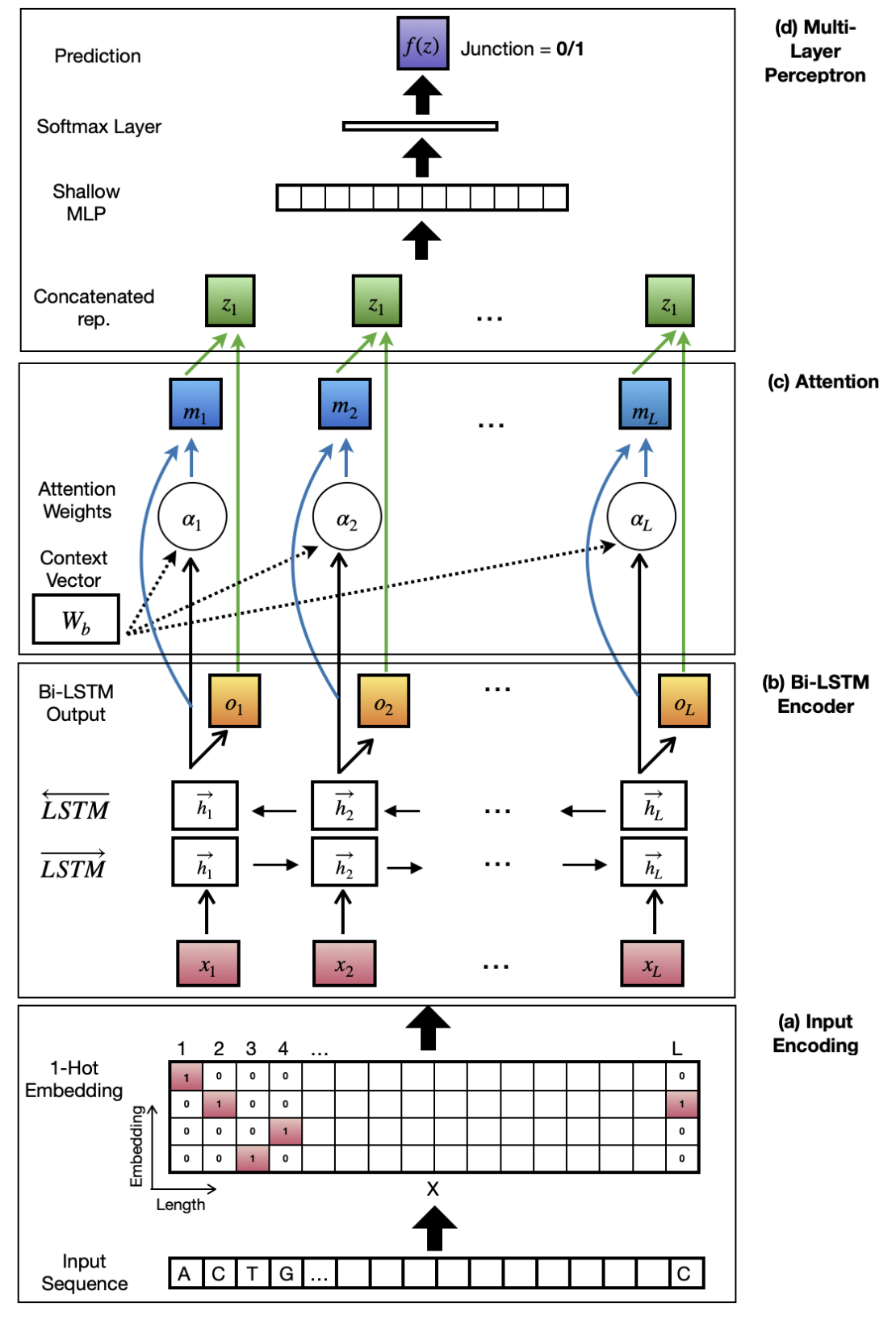}
\caption{The proposed DeepDeCode model architecture. It includes four components: (a) Input encoding; (b) Bi-LSTM encoder; (c) Attention Layer; (d) Multi-layer Perceptron}
\label{fig:model_architecture}
\end{center}
\end{figure}

DeepDeCode consists of four main modules as shown in Figure \ref{fig:model_architecture}. Since the entire model, including the attention mechanisms
is differentiable, we can train the model end-to-end by using backpropagation. We describe the details of each component in Figure \ref{fig:model_architecture}.

\textbf{1-hot encoding}: 
As shown in Figure \ref{fig:model_architecture}, the input DNA sequence of interest ($n^{th}$ pair of the dataset $D$) is denoted as a vector, which is mapped to the matrix $\textbf{X}$ of size $M$ × $L$. 
Here, $M = 4$, which denotes the number of nucleotides in the human genome (they are A, C, T and G). The one-hot encoding is created by putting $1$ for the corresponding position in the embedding for the nucleotide present, and $0$ for the others.
$L$ is the total length of the DNA sequence. The $l^{th}$ column in matrix $\textbf{X}$ is the one-hot encoding vector $\textbf{x}_l$, which represents the signal from the $l^{th}$ nucleotide of the sequence, where $l \in \{1, \dots , L\}$.
Hence, $\textbf{X}$ includes a total of $L$ vector elements that are sequentially ordered along the genome coordinate. 
Considering the sequential nature of such signal reads, we treat each element (essentially a nucleotide position) as a ‘time step’ to be fed into the Bi-LSTM.

\textbf{Bi-LSTM encoder}:  The input matrix $\textbf{X}$  goes into both forward and backward LSTM configured as a Bidirectional LSTM. Since DNA sequences can be modeled from two directions, considering directionality of the sequences would enhance the performance of splice site prediction.
 It includes a forward $\overrightarrow{LSTM}$ that models $\textbf{X}$ from $\textbf{x}_1$ to $\textbf{x}_L$ and a backward $\overleftarrow{LSTM}$ that models $\textbf{X}$ from $\textbf{x}_L$ to $\textbf{x}_1$. For each nucleotide position $l$, the two LSTMs learn meaningful features in a supervised manner, each generating a $d$ dimensional vector representation $\overrightarrow{\textbf{h}_l}$ and $\overleftarrow{\textbf{h}_l}$ as in Equation \ref{eqn:lstm hidden}:
\begin{align}
\label{eqn:lstm hidden}
\begin{split}
    \overrightarrow{\textbf{h}_l} = \overrightarrow{LSTM}(\textbf{x}_l) \\
    \overleftarrow{\textbf{h}_l} = \overleftarrow{LSTM}(\textbf{x}_l)
\end{split}
\end{align}
where $\overrightarrow{\textbf{h}_l}, \overleftarrow{\textbf{h}_l} \in \mathbb{R}^{d \cdot u}$. Here, we have $d$ hidden units in each LSTM layer, and $u$ hidden LSTM layers.

Both $\overrightarrow{\textbf{h}_l}$ and $\overleftarrow{\textbf{h}_l}$ are concatenated to obtain the final embedding vector $\textbf{h}_l = \left[\overrightarrow{\textbf{h}_l}, \overleftarrow{\textbf{h}_l}\right] \in \mathbb{R}^{2\cdot d\cdot u}$ representing the learnt features of each splice junction at the $l^{th}$ position.

\textbf{Attention}:
We use ‘soft’ attention as proposed by \cite{show-attend-tell} to recognize those nucleotides in the input DNA sequence which are \textit{most important} for the splice site classification. This further helps us interpret the results of the model.
The encoding $\textbf{h}_l$ obtained from the Bi-LSTM encoder is passed through the non-linearity layer (output gate of Bi-LSTM) to get the predicted output $\boldsymbol{o}_l \in \mathbb{R}^{2\cdot d}$. 
We apply attention to $\boldsymbol{o}_l$ at each nucleotide position $l$, where $1\leq l \leq L$.
This attention mechanism is implemented by learning a weight vector $\alpha \in \mathbb{R}^{L}$. The $l^{th}$ element of $\alpha$ is denoted as:
\begin{equation}
    \alpha_l = \mathrm{softmax}(\boldsymbol o_l) = \frac{\exp(\boldsymbol{W}_b \cdot \boldsymbol{o}_l )}{\sum_{i=1}^{L} \exp(\boldsymbol{W}_b\cdot \boldsymbol{o}_i)}
\end{equation}
 The scalar $\alpha_l$ is computed over the output representations ${\boldsymbol{o}_1, \ldots , \boldsymbol{o}_L}$. The context parameter $\boldsymbol{W}_b \in \mathbb{R}^ {2d}$ is randomly initialized and jointly learned with the other model parameters during training. Here, we use a single attention head. 
 Given the attention weight $\alpha_l$ of each nucleotide position, we can represent the entire DNA sequence as a weighted sum of all its nucleotide embeddings $\boldsymbol{m} \in \mathbb{R}^{2\cdot d}$: 
\begin{equation}
    \boldsymbol{m} = \sum_{l=1}^{L} \alpha_l \cdot \boldsymbol{o}_l
\end{equation}

Essentially, for $l \in \{1, \dots, L\}$, the attention weights $\alpha_l$ tell us the relative importance of the $l^{th}$ output representation $\boldsymbol{o}_l$.
The representation $\boldsymbol{m}$ concatenated with the encoded embedding $\boldsymbol{h}$ gives the overall representation $\boldsymbol{z}$ of the input DNA sequence, $\boldsymbol{z = [m, h]} \in \mathbb{R}^{(2\cdot d + 2\cdot d\cdot u)}$. 

\textbf{Multi-layer perceptron}:
The vector $\boldsymbol{z}$ summarizes the information of the nucleotide positions for the DNA sequence.
We feed it into a classification module $f(\cdot)$ comprising a multi-layer perceptron (MLP) with a final sigmoid activation. This gives us the probability of there being a splice junction in the input DNA sequence.

\section{Experiments}
 
\subsection{Dataset}
The human DNA sequence data from the December 2013 assembly of the human genome (hg38, GRCh38 Genome Reference Consortium Human Reference 38) was obtained from the UCSC Genome Browser \cite{ucsc}. We obtained the location of exons within the DNA sequences from the latest release (Release 34, GRCh38.p13) of GENCODE annotations \cite{gencode} in the Gene Transfer Format (GTF)\footnote{GTF is a file format used to hold information about gene structure}, which contains comprehensive gene annotations on the reference chromosomes in the human genome.

We created DNA sequences of various lengths, i.e., sequences containing different number of nucleotides ($nt$). For each sequence length, we created datasets of two types: (1) containing acceptor sites; (2) containing donor sites. For each dataset type, each positive sequence in $\textbf{X}$ (i.e., class $1$) contains the splice junction at the midpoint of the sequence. This allowed us to create aligned attention maps for all DNA sequences for intrinsic visualization as described in Section~\ref{subsec:intrinsic}. We extracted a portion of the sequence from the center of each intron or exon to generate a negative sequence not containing a splice junction (i.e., class $0$) of length $L$. 
Table \ref{table:dataset details} shows the average number of sequences we obtained over the entire human genome for the acceptor and donor sites respectively. The dataset is imbalanced, since the number of negative (intronic and exonic) sequences is higher than positive sequences (containing a splice site). 
After balancing the data through undersampling, we created the training, validation and test sets containing 70\%, 10\% and 20\% of the data respectively.

\begin{table}[h]
\centering
\caption{Post-processing data numbers showing the average number of DNA sequences, sequences containing splice sites, sequences created from the exonic region and sequences created from the intronic region.}
\vspace{0.25cm}
\begin{tabular}{|c|c|c|c|c|  } 
 \hline
 \textbf{Splice Site Type} & \textbf{\# Sequences} & \textbf{\# Splice Sites} & \textbf{\# Exons} & \textbf{\# Introns} \\ 
  \hline
  
 \textbf{Acceptor} & 282,743 & 72,742 & 135,500 & 74,501 \\ 
 \textbf{Donor} & 274,296 & 72,255 & 128,927 & 73,114  \\ 
 \hline
 
\end{tabular}
\label{table:dataset details}
\end{table}

\subsection{Experimental Set-Up}
 
We compared the DeepDeCode model to two standard baselines: (1) Convolutional Neural Networks (CNN) \cite{splicerover}; (2) Recurrent Neural Networks (RNN) \cite{dna_level_splice_junc}.\footnote{Code available at \url{https://github.com/asmitapoddar/Deep-Learning-DNA-Sequences}} Each model was trained to identify the acceptor and donor site sequences separately. We also used visualization techniques to understand the importance of each nucleotide in the DNA sequence. 

\textbf{Model hyperparameters}: For DeepDeCode, we set the LSTM embedding size $d$ to $32$. 
Therefore, the context vectors is $\boldsymbol{W}_b$ to size $64$. After hyperparameter tuning over the validation set, values for batch size and dropout were set to 32 and 0.2 respectively. All the parameters described in Section \ref{method} and Figure \ref{fig:model_architecture} are learned together to minimize the cross entropy loss function. We do early stopping to prevent model overfitting and use the Adam optimizer \cite{adam} to train DeepDeCode. 
 
 \subsection{Model Performance}
Tables \ref{table: results1} and \ref{table: results2} show the accuracy and F1 score for the acceptor and donor models for each architecture as lengths of the DNA sequences as varied.
We observe that DeepDeCode performs significantly better than both the CNN and RNN baselines.

\begin{table}[h]
\caption{Results obtained for acceptor site dataset.}
\vspace{0.25cm}
\centering
\begin{tabular}{|c|c c|c c|c c|  } 
 \hline
& \multicolumn{2}{c|}{\textbf{CNN}}&\multicolumn{2}{|c|}{\textbf{RNN}} & \multicolumn{2}{c|}{\textbf{DeepDeCode}}  \\ 
 
\textbf{Sequence Length} & \textbf{Accuracy} & \textbf{F1} & \textbf{Accuracy} & \textbf{F1} & \textbf{Accuracy} & \textbf{F1} \\ 
  \hline
 $20 nt$ & 0.670 & 0.636 & 0.780 & 0.778 & 0.929 & 0.924  \\ 
 $40 nt$ & 0.672 & 0.666 & 0.792 & 0.786 & 0.945 & 0.940 \\ 
 $60 nt$& 0.701 &0.705 & 0.746 & 0.754 & 0.948 & 0.953  \\ 
 $80 nt$ & 0.718 & 0.708 & 0.728 & 0.710 & 0.953 & 0.950 \\ 
 $100 nt$ & 0.682 & 0.681 & 0.711 & 0.708 & 0.932 & 0.931 \\ 
 \hline
\end{tabular}
\label{table: results1}
\end{table}

\begin{table}[h]
\caption{Results obtained for donor site dataset.}
\vspace{0.25cm}
\centering
\begin{tabular}{|c|c c|c c|c c|  } 
 \hline
& \multicolumn{2}{c|}{\textbf{CNN}}&\multicolumn{2}{|c|}{\textbf{RNN}} & \multicolumn{2}{c|}{\textbf{DeepDeCode}} \\ 
 
\textbf{Sequence Length} & \textbf{Accuracy} & \textbf{F1} & \textbf{Accuracy} & \textbf{F1} & \textbf{Accuracy} & \textbf{F1} \\ 
  \hline
 $20 nt$ & 0.657 & 0.626 & 0.761 & 0.732  & 0.939 & 0.933  \\ 
 $40 nt$& 0.662 & 0.646 & 0.766 & 0.786 & 0.949 & 0.947 \\ 
 $60 nt$& 0.683 & 0.655 & 0.738 & 0.759 & 0.956 & 0.950  \\ 
 $80 nt$ & 0.689 & 0.657 & 0.720 & 0.710 & 0.922 & 0.922  \\ 
 $100 nt$ & 0.685 & 0.641 & 0.710 & 0.689 & 0.916 & 0.904 \\ 
 \hline
\end{tabular}

\label{table: results2}
\end{table}

We also observe that the performance increases as
the length of the input sequences increases, until the performance peaks at sequence length of $80 nt$ for acceptor site sequences, and $60 nt$ for donor site sequences. This suggests that as the sequence length increases upto these lengths for the respective splice sites, the models can model the dependencies among the nucleotides to give improved results. Both CNN and RNN-based models are unable to learn very long-term dependencies.

\section{Visualization}
 
We visualize the model prediction using two methods --- intrinsic and extrinsic visualization. In this section, we show the two types of visualizations across the acceptor dataset for sequences of length $L = 80 nt$. Visualizations across the donor dataset can be found in the Appendix \ref{appendix}.

To visualize the actual nucleotides, we use sequence logos.
Figure \ref{fig:visualisation} (I) shows the sequence logos for the dataset.
The probability $p^{m}_l$ of each nucleotide (letter) $m$ occurring at a particular position $l$ in the DNA sequence is computed as the count $m_l$ of the sequences containing the letter $m$ occurring at the position $l$ over all the $n$ sequences in the dataset:
\begin{equation}
    p^{m}_l = \frac{m_l}{n}
\end{equation}
where $m \in M; M= \{A, C, T, G\}$.
The size of each letter is made proportional to its probability, and the letters are
sorted so the most probable one is at the bottom. Hence, we can determine the information content at every position in the sequence.

 \subsection{Intrinsic Visualization --- Attention Maps for Pattern Interpretation} \label{subsec:intrinsic}
 
 Intrinsic interpretability can be realized by designing self-explanatory models which incorporate interpretability directly into the model structures \cite{interpretable-ml}. The attention weights $\alpha$ learned by DeepDeCode inform the importance of each nucleotide position in predicting whether a particular sequence contains a splice site. Thus, the information conveyed by attention maps can be used to make biological inferences. To get the attention maps, we compute the average attention values across all nucleotide positions for all the sequences containing a splice site in the training dataset. 

\begin{figure}[h!]
\centering
\includegraphics[width=0.95\textwidth]{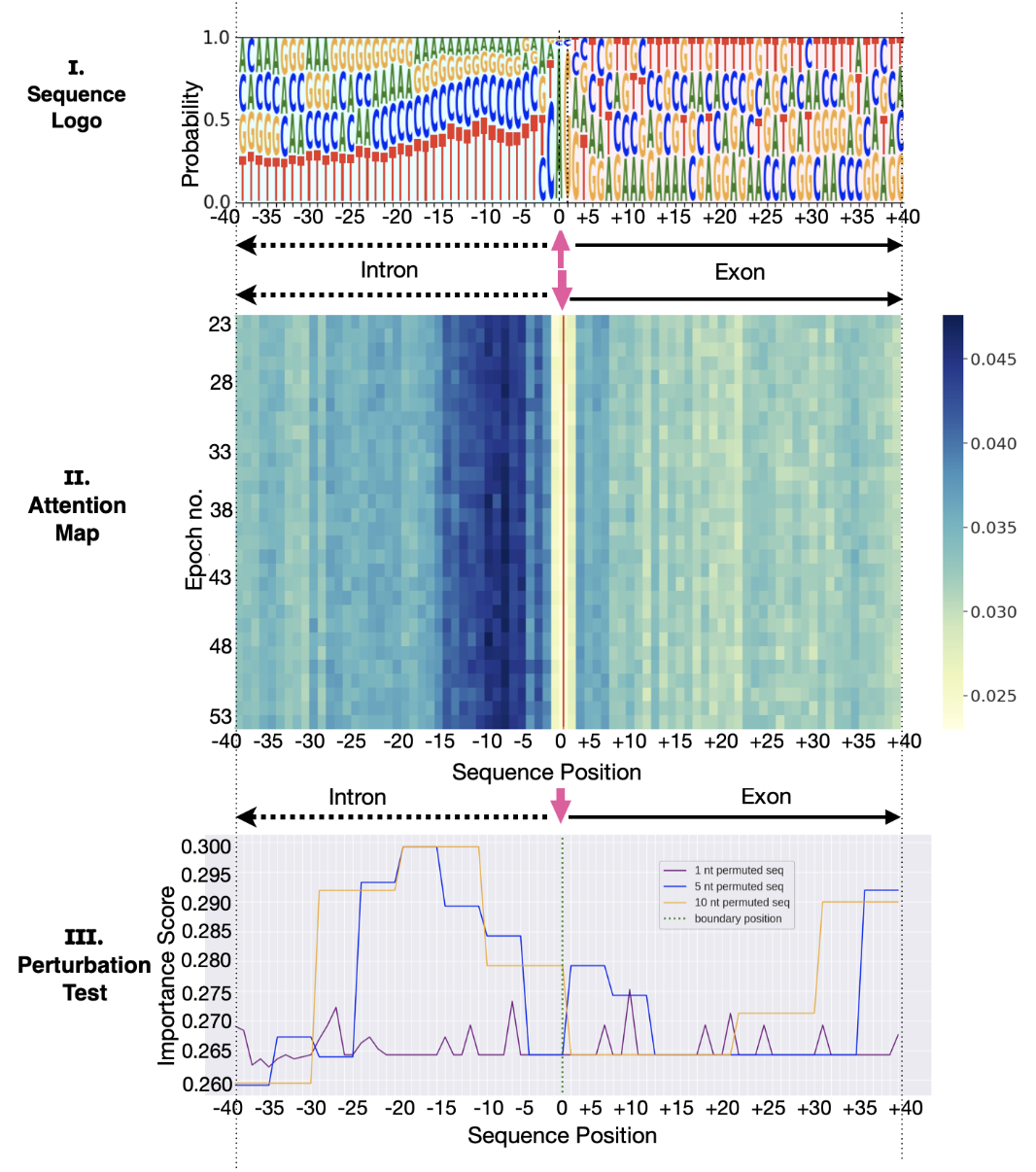}
\caption{Visualization of acceptor site dataset over the DNA sequences of length $80$ nucleotides ($nt$). Figure \ref{fig:visualisation}(I) shows the sequence logo, Figure \ref{fig:visualisation}(II) shows attention maps and Figure \ref{fig:visualisation}(III) shows results of the perturbation test for the DNA sequence. The $x$ axis denotes the position of a nucleotide ($nt$) in the sequence, where position $0$ denotes the splice junction.
} 
\label{fig:visualisation}
\end{figure}

In Figure \ref{fig:visualisation} (II), we show the attention heatmaps over the DNA sequence for the last 30 epochs of network training. This visualizes how the distribution of the attention weights develops during the training process as the network learns to pay attention.

It is well known in the literature \cite{mo_biology}, that due to the process of splicing, the Polypyrimidine (PY) Tract is an important region of the DNA sequence. The $PY$-Tract, which is usually $5$-$15 nt$ long, is located in the region $[-35, -5] nt$. As shown by the sequence logo, this region of the sequence is typically represented by the nucleotide pattern $CT$. 
From Figure \ref{fig:visualisation}(II), we observe that DeepDeCode focuses on the above region in the DNA sequence. 
 This shows that it can identify correct regions and sequence motifs for determining the acceptor sites.
More analysis about the attention maps for donor sites can be found in the Appendix \ref{appendix}.

 \subsection{External Visualization --- Sequence Perturbation tests}
External visualization techniques aim to provide an understanding about the knowledge that a trained model has already acquired \cite{interpretable-ml}. It tries to answer the question: ``\textit{which parts of the input, if they were not seen by the model, would most change its prediction?}" For each DNA sequence in the test set, we perform computational sequence perturbations to evaluate the effect of mutating every nucleotide of an input sub-sequence on the neural network model's accuracy. Such procedures are similar to the site-specific mutagenesis \cite{mutagenesis} performed in a wet lab. 

Given a dataset of DNA sequences of length $L$, for a fixed window $r=[x,y]$ of nucleotide positions; the perturbation technique is applied to obtain an importance score $s_r$ of the nucleotide(s):
\begin{equation}
    s_r = a^{M}_{orig} - a^{M}_{mut}
\end{equation}
where $a^{M}_{orig}$ is the accuracy for the trained model $M$ on the dataset of original sequences, and $a^{M}_{mut}$ is the test accuracy for model $M$ on the dataset of sequences mutated in the nucleotide range $r=[x,y]$; such that $x,y \in \{1,\dots,L\}$, and $x<y$ (i.e., the range $r$ is within the sequence).

The importance score $s_r$ of a genomic region reflects the contribution of the nucleotide pattern(s) in that region to the classification task. The mutated sequences are created by destroying information contained in the region $r$ under study by replacing the one-hot encoding of the nucleotide(s) under consideration with Gaussian noise. We select the range of consecutive nucleotides $r$ to be permuted at a time, in the following ways: $1 nt$ input perturbation, $5 nt$ input perturbation, $10 nt$ input perturbation.

Figure \ref{fig:visualisation}(III) shows how the model accuracy changes with input perturbation. We observe that high attention regions have higher importance scores. This further confirms the reliability of the attention values obtained by DeepDeCode.
We do not observe much variation in the importance scores over the sequence on mutating $1 nt$ of the sequence at a time. This is because most of the crucial motifs in the DNA sequence are not singular, and no conclusions about mutations can be drawn from this sequence perturbation test.

As we mutate $5 nt$ or $10 nt$ of the sequence at a time, we have a higher probability of mutating a PY-Tract (which is usually $5$-$15 nt$ long in the intronic region). We note a significant difference in accuracy in the $10 nt$ perturbation test as more information gets destroyed in the relevant regions. For $5 nt$ and $10 nt$ perturbation tests, we observe that the regions of the sequence with high attention values have higher perturbation importance scores, confirming that these regions contain important motif information for the model to perform well for the classification task.

Hence, DeepDeCode can help us understand and visualize the difference in salient signals and splicing grammar at the beginning and end of exons. 
Our visualization methods build trust in the deep learning models by validating biological information that have been found by wet lab methods.
Our method not only confirms known genomic information, but can also be used in discovering new motifs and patterns informing splice junctions that were not previously known.

\section{Conclusion}

Our proposed deep learning architecture DeepDeCode performs significantly better than other baselines, achieving  $95\%$ accuracy in distinguishing between sequences containing splice junctions and those that do not. To build trust in our methods, we generate attention maps of DeepDeCode over the sequences, and corroborate the intrinsic visualization by performing sequence perturbation test, an extrinsic visualization technique. We conceptualize the sequence motifs identified through sequence logos, and see that our model can determine the biologically important regions while making predictions. Through our visualization method, our model can therefore help highlight further areas of interest in the genome and be used to make inferences on yet undiscovered biological phenomena.
 Given the results of our methodology, we expect that such approaches can be used to help to battle cancer and pilot remedies for other diseases that are caused due to genomic discrepancies.

Future research endeavours can involve building models that can predict both acceptor and donor splice sites simultaneously---this can help to identify entire exons in the genome. Our method can also be applied to other genome sequencing challenges such as promoter region recognition \cite{promid}, predicting enhancers \cite{enhancer} and the study of epigenetic marks for gene expression \cite{DNA-methylation-data}. The visualisation techniques that we have developed in this paper can further be used for building trustworthiness when solving these tasks. 

\bibliography{refs}
\bibliographystyle{plain}

\appendix
\section{DeepDeCode for DNA sequences containing donor sites} \label{appendix}
\begin{figure}[ht]
\centering
\includegraphics[width=0.95\textwidth]{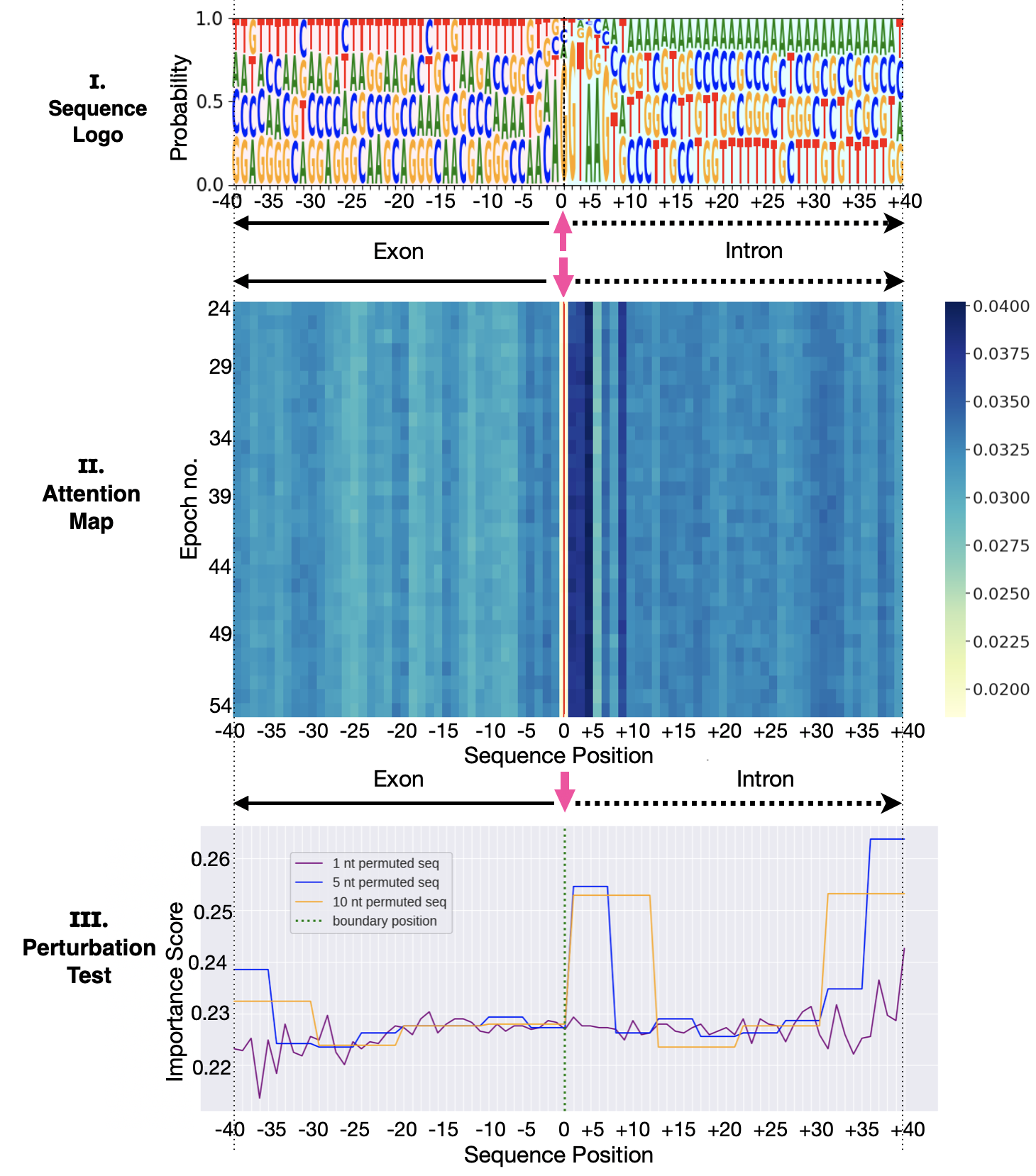}
\caption{Visualization of donor site dataset over the DNA sequences of length $80$ nucleotides ($nt$). Figure \ref{fig:visualisation_donor}(I) shows the sequence logo, Figure \ref{fig:visualisation_donor}(II) shows attention maps and Figure \ref{fig:visualisation_donor}(III) shows results of the perturbation test for the DNA sequence. The $x$ axis denotes the position of a nucleotide in the sequence, where position $0$ denotes the splice junction. For donor sequences, the exons occur downstream of the splice site, and vice versa for introns. } 
\label{fig:visualisation_donor}
\end{figure}

This section contains the results of DeepDeCode applied to DNA sequences containing donor sites. 
Figure \ref{fig:visualisation_donor} shows the sequence logos, attention maps and result of the perturbation test.

For the donor sites, most of the DeepDeCode attention is on the intronic region just after the splice site, i.e., $[+1, +2] nt$ upstream (in the intronic region of the sequences), which is rich in the nucleotides $GT$ (containing $G$ at position $+1$ and $T$ at position $+2$). It is observed that some attention is also paid to in the region slightly further upstream of the splice site ($[+5, +7] nt$ upstream). This is due to spliceosomal twin introns (stwintrons), where the $GT$ sequences involved in splicing are interrupted by another intron (internal intron) \cite{swintron}. There are alternative motifs ($GT$) repeated $3 nt$ apart from each other within the intron.
Hence, we observe a wobbling effect in the attention maps of DeepDeCode, with attention being paid at the positions $[+1, +2] nt$ and $[+5, +7] nt$ due to the the dual canonical motifs $GT$ separated by $3nt$, in the intronic region. 
The sequence perturbation test also shows that we get a higher importance score in the above mentioned regions.

\end{document}

%% file: math_commands.tex
\usepackage{amsmath,amsfonts,bm}

\def\eqref#1{equation~\ref{#1}}

\def\1{\bm{1}}

\DeclareMathAlphabet{\mathsfit}{\encodingdefault}{\sfdefault}{m}{sl}
\SetMathAlphabet{\mathsfit}{bold}{\encodingdefault}{\sfdefault}{bx}{n}













%% file: main.bbl
\begin{thebibliography}{10}

\bibitem{gene-regulation}
Babak Alipanahi, Andrew Delong, Matthew~T Weirauch, and Brendan~J Frey.
\newblock Predicting the sequence specificities of dna-and rna-binding proteins by deep learning.
\newblock {\em Nature biotechnology}, 33(8):831--838, 2015.

\bibitem{protein-struct-pred}
Ehsaneddin Asgari, Nina Poerner, Alice McHardy, and Mohammad Mofrad.
\newblock Deepprime2sec: Deep learning for protein secondary structure prediction from the primary sequences.
\newblock {\em bioRxiv}, page 705426, 2019.

\bibitem{bioinformatics}
Ardeshir Bayat.
\newblock {Clinical review Science, medicine, and the future Bioinformatics}.
\newblock {\em British Medical Journal}, 324(April):1018--1022, 2002.

\bibitem{DNA-methylation-data}
Christoph Bock.
\newblock Analysing and interpreting dna methylation data.
\newblock {\em Nature Reviews Genetics}, 13(10):705--719, 2012.

\bibitem{bert}
Jacob Devlin, Ming-Wei Chang, Kenton Lee, and Kristina Toutanova.
\newblock Bert: Pre-training of deep bidirectional transformers for language understanding.
\newblock {\em arXiv preprint arXiv:1810.04805}, 2018.

\bibitem{interpretability}
K.K. Dey, B.~van~de Geijn, S.S. Kim, and et~al.
\newblock Evaluating the informativeness of deep learning annotations for human complex diseases.
\newblock {\em Nature Communications}, 11(4703), 2020.

\bibitem{interpretable-ml}
Mengnan Du, Ninghao Liu, and Xia Hu.
\newblock {Techniques for interpretable machine learning}.
\newblock {\em Communications of the ACM}, 63(1):68--77, 2020.

\bibitem{splicevec}
Aparajita Dutta, Tushar Dubey, Kusum~Kumari Singh, and Ashish Anand.
\newblock Splicevec: distributed feature representations for splice junction prediction.
\newblock {\em Computational biology and chemistry}, 74:434--441, 2018.

\bibitem{genelinguistics}
Gemma~Bel Enguix and M~{Dolores Jim{\'{e}}nez-L{\'{o}}pez}.
\newblock {Natural Language And The Genetic Code: From The Semiotic Analogy To Biolinguistics}.
\newblock {\em Proceedings of the 10th World Congress of the International Association for Semiotic Studies (IASS/AIS)}, pages 771--780, 2012.

\bibitem{genomics-scope}
Brian~S. Gloss and Marcel~E. Dinger.
\newblock {Realizing the significance of noncoding functionality in clinical genomics}.
\newblock {\em Experimental and Molecular Medicine}, 50(8), 2018.

\bibitem{gencode}
Jennifer Harrow, Adam Frankish, Jose~M. Gonzalez, Electra Tapanari, Mark Diekhans, Felix Kokocinski, Bronwen~L. Aken, Daniel Barrell, Amonida Zadissa, Stephen Searle, If~Barnes, Alexandra Bignell, Veronika Boychenko, Toby Hunt, Mike Kay, Gaurab Mukherjee, Jeena Rajan, Gloria Despacio-Reyes, Gary Saunders, Charles Steward, Rachel Harte, Michael Lin, C{\'{e}}dric Howald, Andrea Tanzer, Thomas Derrien, Jacqueline Chrast, Nathalie Walters, Suganthi Balasubramanian, Baikang Pei, Michael Tress, Jose~Manuel Rodriguez, Iakes Ezkurdia, Jeltje {Van Baren}, Michael Brent, David Haussler, Manolis Kellis, Alfonso Valencia, and Alexandre Reymond.
\newblock {GENCODE: The reference human genome annotation for the ENCODE project}.
\newblock {\em Genome Research}, 22(9):1760--1774, 2012.

\bibitem{lstm}
Sepp Hochreiter and J{\"u}rgen Schmidhuber.
\newblock Long short-term memory.
\newblock {\em Neural Computation}, 9(8):17351780, 1997.

\bibitem{primary-sequence-splicing-dl}
Kishore Jaganathan, Sofia~Kyriazopoulou Panagiotopoulou, Jeremy~F McRae, Siavash~Fazel Darbandi, David Knowles, Yang~I Li, Jack~A Kosmicki, Juan Arbelaez, Wenwu Cui, Grace~B Schwartz, et~al.
\newblock Predicting splicing from primary sequence with deep learning.
\newblock {\em Cell}, 176(3):535--548, 2019.

\bibitem{DNABERT}
Yanrong Ji, Zhihan Zhou, Han Liu, and Ramana~V Davuluri.
\newblock {DNABERT: pre-trained Bidirectional Encoder Representations from Transformers model for DNA-language in genome}.
\newblock {\em Bioinformatics}, 37(15):2112--2120, 02 2021.

\bibitem{swintron}
Napsug{\'a}r Kavalecz, Norbert {\'A}g, Levente Karaffa, Claudio Scazzocchio, Michel Flipphi, and Erzs{\'e}bet Fekete.
\newblock A spliceosomal twin intron (stwintron) participates in both exon skipping and evolutionary exon loss.
\newblock {\em Scientific reports}, 9(1):1--11, 2019.

\bibitem{ucsc}
W~James Kent, Charles~W Sugnet, Terrence~S Furey, Krishna~M Roskin, Tom~H Pringle, Alan~M Zahler, and David Haussler.
\newblock The human genome browser at ucsc.
\newblock {\em Genome research}, 12(6):996--1006, 2002.

\bibitem{adam}
Diederik~P Kingma and Jimmy Ba.
\newblock Adam: A method for stochastic optimization.
\newblock {\em arXiv preprint arXiv:1412.6980}, 2014.

\bibitem{mutagenesis}
Thomas~A Kunkel.
\newblock Rapid and efficient site-specific mutagenesis without phenotypic selection.
\newblock {\em Proceedings of the National Academy of Sciences}, 82(2):488--492, 1985.

\bibitem{dna_level_splice_junc}
Byunghan Lee, Taehoon Lee, Byunggook Na, and Sungroh Yoon.
\newblock Dna-level splice junction prediction using deep recurrent neural networks.
\newblock {\em arXiv preprint arXiv:1512.05135}, 2015.

\bibitem{mo_biology}
Harvey Lodish, Arnold Berk, S~Lawrence Zipursky, Matsudaira Paul, Baltimore David, and Darnell James.
\newblock {\em {Molecular Cell Biology, 4th edition}}.
\newblock New York: W. H. Freeman, 2000.

\bibitem{danq}
Daniel Quang and Xiaohui Xie.
\newblock Danq: a hybrid convolutional and recurrent deep neural network for quantifying the function of dna sequences.
\newblock {\em Nucleic acids research}, 44(11):e107--e107, 2016.

\bibitem{highthroughputseq}
Jason~A Reuter, Damek~V Spacek, and Michael~P Snyder.
\newblock {High-throughput sequencing technologies}.
\newblock {\em Physiology {\&} behavior}, 176(12):139--148, 2015.

\bibitem{bilstm-rnn}
Noopur Singh, Ravindra~Nath Katiyar, and Dev~Bukhsh Singh.
\newblock {Splice-Site Identification for Exon Prediction Using Bidirectional Lstm-Rnn Approach}.
\newblock {\em Biochemistry and Biophysics Reports}, 2022.

\bibitem{attend&predict}
Ritambhara Singh, Jack Lanchantin, Arshdeep Sekhon, and Yanjun Qi.
\newblock {Attend and predict: Understanding gene regulation by selective attention on chromatin}.
\newblock {\em Advances in Neural Information Processing Systems}, 2017.

\bibitem{enhancer}
Asa Thibodeau, Asli Uyar, Shubham Khetan, Michael~L Stitzel, and Duygu Ucar.
\newblock A neural network based model effectively predicts enhancers from clinical atac-seq samples.
\newblock {\em Scientific reports}, 8(1):1--15, 2018.

\bibitem{promid}
Ramzan Umarov, Hiroyuki Kuwahara, Yu~Li, Xin Gao, and Victor Solovyev.
\newblock Promoter analysis and prediction in the human genome using sequence-based deep learning models.
\newblock {\em Bioinformatics}, 35(16):2730--2737, 2019.

\bibitem{splicefinder}
Ruohan Wang, Zishuai Wang, Jianping Wang, and Shuaicheng Li.
\newblock Splicefinder: ab initio prediction of splice sites using convolutional neural network.
\newblock {\em BMC bioinformatics}, 20(23):652, 2019.

\bibitem{pierre-robin}
Jessie~X. Xu, Nicky Kilpatrick, Naomi~L. Baker, Anthony Penington, Peter~G. Farlie, and Tiong~Yang Tan.
\newblock {Clinical and molecular characterisation of children with pierre robin sequence and additional anomalies}.
\newblock {\em Molecular Syndromology}, 7(6):322--328, 2016.

\bibitem{show-attend-tell}
Kelvin Xu, Jimmy Ba, Ryan Kiros, Kyunghyun Cho, Aaron Courville, Ruslan Salakhudinov, Rich Zemel, and Yoshua Bengio.
\newblock Show, attend and tell: Neural image caption generation with visual attention.
\newblock In {\em International conference on machine learning}, pages 2048--2057, 2015.

\bibitem{pangolin}
T.~Zeng and Y.I. Li.
\newblock Predicting rna splicing from dna sequence using pangolin.
\newblock {\em Genome Biology}, 23(103), 2022.

\bibitem{cancer}
Wei Zhang, Ana Bojorquez-Gomez, Daniel~Ortiz Velez, Guorong Xu, Kyle~S. Sanchez, John~Paul Shen, Kevin Chen, Katherine Licon, Collin Melton, Katrina~M. Olson, and et~al.
\newblock A global transcriptional network connecting noncoding mutations to changes in tumor gene expression.
\newblock {\em Nature News}, Apr 2018.

\bibitem{deepsplice}
Yi~Zhang, Xinan Liu, James~N. Macleod, and Jinze Liu.
\newblock {DeepSplice: Deep classification of novel splice junctions revealed by RNA-seq}.
\newblock {\em Proceedings - 2016 IEEE International Conference on Bioinformatics and Biomedicine, BIBM 2016}, pages 330--333, 2017.

\bibitem{splicerover}
Jasper Zuallaert, Fr{\'{e}}deric Godin, Mijung Kim, Arne Soete, Yvan Saeys, and Wesley {De Neve}.
\newblock {Splicerover: Interpretable convolutional neural networks for improved splice site prediction}.
\newblock {\em Bioinformatics}, 34(24):4180--4188, 2018.

\end{thebibliography}
